%% file: PRL_LED_final.tex
\documentclass[aps,prl,twocolumn,showpacs,groupedaddress,fixfloat]{revtex4}  
\usepackage{graphicx}  
\usepackage{dcolumn}   
\usepackage{bm}        
\usepackage{amssymb}   
\input epsf.sty

\def\ppbar{$p\overline{p} $}            
\def\met{\mbox{${\hbox{$E$\kern-0.6em\lower-.1ex\hbox{/}}}_T$}} 
\def\ipb{pb$^{-1}$}                     
\def\D0{D\O}                            

\def\ct{\mbox{$\cos\theta^*$}}
\def\pt{$p_{T}$}

\def\gkk{$G_{\rm KK}$}

\begin{document}



\title{Search for Large Extra Spatial Dimensions in Dimuon Production with the \D0 Detector}
\input list_of_authors_r2.tex  
\date{August 19, 2005}

\begin{abstract}
We present the results of a search for the effects of large extra spatial dimensions in $p{\bar p}$ collisions at $\sqrt{s} =$ 1.96 TeV in events containing a pair of energetic muons.  The data correspond to 246 \ipb\ of integrated luminosity collected by the \D0\ experiment at the Fermilab Tevatron Collider. Good agreement with the expected background was found, yielding no evidence for large extra dimensions.  We set 95\% C.L. lower limits on the fundamental Planck scale between 0.85~TeV and 1.27~TeV within several formalisms. These are the most stringent limits achieved in the dimuon channel to date.
\end{abstract}

\maketitle
In their 1998 paper, Arkani-Hamed, Dimopoulos, and Dvali (ADD) suggested that the seemingly unreachable Planck energy scale (conventionally thought to be $M_{\rm Pl} \sim 10^{19}$ GeV) may be in fact much lower, i.e. within the reach of current and planned future colliders~\cite{ADD}.  They postulated that the standard model (SM) particles and gauge interactions are confined to a three-dimensional ``brane'' embedded in a ``multiverse'', which consists of the three standard plus $n$ additional compact spatial dimensions.  However, gravitons in this framework can propagate in the entire multiverse. The gravitons propagating in compact extra dimensions appear as a tower of Kaluza-Klein (KK) excited modes from the point of view of the SM brane. Furthermore, the radius of compactification ($R$) of extra dimensions in the ADD model is much larger than either the Planck or electroweak length, and may be as large as $\sim 1$ mm. Since gravitons are free to propagate in these large extra dimensions, the gravitational interaction would appear suppressed on the SM brane, due to the extra volume gravity permeates. Consequently, while the apparent Planck scale is $\sim 10^{19}$ GeV, with respect to the $3+n$-dimensional space, the fundamental Planck scale ($M_{S}$) can be as low as $\sim 1$ TeV, thus eliminating the hierarchy problem of the SM.  

\indent	The phenomenological consequences of the ADD model have been a subject of intense study in recent years.  For a review of the possible effects of large extra dimensions, ranging from modification of Newtonian gravity to black hole production at
future colliders, see, e.g., Ref.~\cite{review1,review2}.

\indent	In this Letter, we describe a search for the effects of large extra dimensions via virtual Kaluza-Klein graviton (\gkk) exchange in $p\bar p$ collisions resulting in the dimuon final state.  Technically, virtual graviton effects are sensitive to the ultraviolet cutoff required to keep the divergent sum over the KK states finite~\protect\cite{GRW,HLZ,Hewett}, rather than the fundamental Planck scale. As the two scales are expected to be closely related, we do not distinguish between them in this analysis.  The search is based on $246\pm16$~pb$^{-1}$ of data collected in 2002--2004 by the D\O\ detector operating at the Fermilab Tevatron Collider at $\sqrt{s} = 1.96$~TeV. We used the method of Refs.~\cite{KCGL,LEDPRL}, in which the dilepton invariant mass ($M$) and the cosine of the scattering angle (\ct) in the dilepton center of mass frame are analyzed simultaneously for the effects of large extra dimensions. This is the first search for large extra dimensions in the dimuon channel at a hadron collider. Previous searches for virtual graviton effects in various channels at LEP, HERA, and the Tevatron are reviewed in detail in Ref.~\cite{review2}.

The dimuon production cross section in the presence of extra dimensions is given by~\cite{GRW,HLZ,Hewett}:
\begin{eqnarray}
	\frac{d^2\sigma}{dMd\ct} & = & f_{\rm SM} + f_{\rm int}\; \eta_G + f_{\rm KK}\; \eta_G^2, \label{eq4}
\end{eqnarray}
where $f_{\rm SM}$, $f_{\rm int}$, and $f_{\rm KK}$ are functions of $(M,\ct)$ and denote the SM, interference, and $G_{\rm KK}$ terms. The effects of large extra dimensions are parametrized via a single variable $\eta_G = {\cal F}/M_S^4$, where ${\cal F}$ is a dimensionless parameter of order unity.  Three different formalisms for ${\cal F}$ are explored in this analysis:
\begin{eqnarray}
	{\cal F} & = & 1 \label{eq1} 
\end{eqnarray}
[Giudice-Rattazzi-Wells (GRW)~\cite{GRW}];
\begin{eqnarray}	
	{\cal F} & = & \left\{ \begin{array}{ll} 
         \log\left( \frac{M_S^2}{M^2} \right), & n = 2 \\
	   \frac{2}{n-2}, & n > 2
	   \end{array} \right. \label{eq2}
\end{eqnarray}
[Han-Lykken-Zhang (HLZ)~\cite{HLZ}];
\begin{eqnarray}	
	{\cal F} & = & \frac{2\lambda}{\pi} = \pm\frac{2}{\pi} \label{eq3}
\end{eqnarray}
(Hewett~\cite{Hewett}).  In Eq.~(\ref{eq3}) $\lambda=\pm1$ indicates whether virtual graviton exchange interferes constructively or destructively with SM processes.  While virtual graviton exchange does not depend strongly on $n$ (the number of extra dimensions), the HLZ formalism for ${\cal F}$ does explore this dependence explicitly.

The D\O\ detector and its data acquisition system are described in detail elsewhere~\cite{D0}. Here we give a brief description of the components used in the analysis.  At the center of the \D0\ detector is the central-tracking system, which consists of a silicon microstrip tracker and a central fiber tracker. Both trackers are located within a 2~T axial magnetic field~\cite{D0}, with designs optimized for tracking and vertexing at pseudorapidities $|\eta_{d}|<3$.  The pseudorapidity $\eta$ is defined as $-\ln\left(\tan\frac{\theta}{2}\right)$, where $\theta$ is the polar angle w.r.t. the proton-beam direction, as measured from the interaction vertex. We also define $\eta_d$, which is the pseudorapidity based on the polar angle measured from the geometric center of the detector. The nearly hermetic uranium/liquid-argon calorimeter is used to measure energies of electrons, photons, and hadrons~\cite{run1det}. The muon system covers $|\eta_{d}|<2$ and consists of a layer of tracking detectors and scintillation trigger counters in front of 1.8~T iron toroidal magnets, followed by two more similar layers of detectors outside the toroids~\cite{muonsys}.  Luminosity is measured using plastic scintillator arrays placed in the large $|\eta_{d}|$ (forward) regions of the detector. 

The sample of candidate events used in the search was collected with a set of triggers that require either one or two muon candidates in the muon system. After detailed event reconstruction an event must contain at least two muon candidates, each matched with a track in the central tracker, which is used for muon momentum measurement. The matching tracks were required to have transverse momenta $p_T > 15$ GeV, at least one hit in the silicon microstrip tracker, and at least nine hits in the central fiber tracker. The latter two requirements ensure reliable momentum measurement, especially at high $p_T$.
To reduce background from cosmic rays, we introduced additional criteria. Since cosmic muons are not correlated with the beam crossing, we required the muon arrival time, as measured in the muon system scintillation counters, to be within $10$ ns (four standard deviations) of the expected arrival time for an highly-relativistic particle produced in a $p\bar p$ collision in the center of the detector~\cite{muonsys}.  Furthermore, dimuon events that originate from a cosmic muon are back-to-back in $\eta$.  This is because the same cosmic muon is reconstructed twice in the event, once when entering the detector and once when leaving it.  Consequently, dimuon events from cosmic rays should have $\eta_1+\eta_2 \approx 0$.  In true dimuon events originating from \ppbar\ collisions, the two muons are generally not back-to-back in $\eta$ due to a longitudinal boost of the dimuon system.  Therefore the sum $\eta_1+\eta_2$ was required to be away from zero by at least five standard deviations of the Gaussian distribution as observed in a cosmic ray sample. The signal efficiency for this selection is ($99\pm1$)$\%$.  After the above selections the cosmic muon contamination in the candidate sample is negligible.  \\
\indent Muons from graviton decay are expected to be isolated from other energetic particles or jets. In contrast, high \pt\ muons from b and c quark decays tend to be non-isolated.  To reduce this copious background, we required that each muon have: (i) $\sum_{{\cal R} = 0.5}(p_T) < 2.5$ GeV, where $\sum_{{\cal R} = 0.5}(p_T)$ is the scalar sum of the $p_{T}$ of all additional tracks contained within a cone of radius ${\cal R} = \sqrt{(\Delta\eta)^2 + (\Delta\phi)^2} = 0.5$~\cite{phi} centered on the muon track; and (ii) $\sum_{{\cal R} = 0.4}(E_T)-\sum_{{\cal R} = 0.1}(E_T) < 2.5$~GeV, where the $E_T$'s are the transverse energies of the calorimeter cells within the respective cones centered on the muon track.  \\
\indent After all the selections are applied the efficiency per muon is $(80\pm 4)\%$, as measured with $Z \to \mu\mu$ events.  This includes the efficiency for muon and track reconstruction, track matching, number of tracker hits, and cosmic ray muon vetos, as well as the isolation selections.  \\
\indent Because this analysis focuses on very high-$p_T$ objects, the most problematic background is that from mismeasured Drell-Yan (DY) events that appear to have very high mass.  This occurs because if a muon's $p_T$ is mismeasured toward a higher value it tends to be grossly mismeasured high, and thus the reconstructed mass also tends to be much higher than it actually is.  This is due to the resolution of the central tracker, which is approximately Gaussian in $1/p_T$ with a typical $\sigma=0.00272$.  We reduce the effect by scaling the $p_{T}$ of each muon track to a weighted average based on the original track $\frac{1}{p_{T}}$ measurement and its uncertainty.  For example, in the highest mass event the original muon $p_{T}$'s were 250 GeV and 1000 GeV, while after $p_{T}$-fixing both $p_{T}$'s became 400 GeV.  This procedure assumes the two muons' transverse  momenta should be equal.  For high mass objects such as a KK graviton the equal momenta assumption would be accurate.  The following equation illustrates the procedure:
\begin{equation}
      \frac{1}{p'_{T_{1}}}= \frac{1}{p'_{T_{2}}}= \frac{|w_{1}/p_{T_{1}} + \epsilon w_{2}/p_{T_{2}}|}{w_{1}+w_{2}},
\end{equation} where $p_{T_{1,2}}$ are the original $p_T$'s of the two muons, $w_{1,2}=1/\sigma^{2}(1/p_{T_{1,2}})$ are the Gaussian weights, and $\sigma(1/p_{T_{1,2}})$ is the uncertainty on $1/p_{T_{1,2}}$ measured from the shape of the $Z$ boson peak.  In the equation above $\epsilon = +1$ for muon pairs with opposite charge and $\epsilon = -1$ for same charge muon pairs.  This search does not require that the two muons have opposite charges, because the efficiency of such a requirement degrades quickly at high masses and the requirement does not reduce the already low background there.

\indent Based on the values of $1/p'_{T}$ and the original angular information, a new momentum was calculated for the two muons in the event.  The same procedure was also applied to the simulated signal and background. After the reweighting was done we required the dimuon mass $M$ to be greater than $50$ GeV, which resulted in a final candidate sample of 17,128 events, most of them in the vicinity of the $Z$ boson peak.  

We modeled the effects of large extra dimensions via the parton-level leading-order (LO) Monte Carlo (MC) generator of Ref.~\cite{KCGL}, augmented with a parametric simulation of the D\O\ detector.  The simulation takes into account detector acceptance and resolution for muons.  The generator includes effects of initial state radiation as described below, and different parton distribution functions (PDF). We used the leading order CTEQ5L~\cite{CTEQ} PDF to estimate the nominal prediction. The parameters of the detector model were tuned using $Z \to \mu\mu$ events. The simulation includes SM DY contributions ($Z/\gamma^*$), Kaluza-Klein graviton exchange diagrams, and their interference.

Since the MC generator contains only the LO parton-level processes, we modeled next-to-leading order (NLO) effects by adding a transverse momentum to the dimuon system.  The model is based on the transverse momentum spectrum of dielectron candidates observed in the calorimeter, which provides better high energy resolution capability than that of the central tracking system.  Since the parton-level cross section is calculated at LO, we accounted for the NLO enhancement in the SM background by scaling the cross sections by a constant $K$-factor of 1.34~\cite{K-factor}.  We assigned a $\pm5\%$ systematic uncertainty on the value of the $K$-factor to account for its mass dependence. We assumed the same constant $K$-factor for the extra dimensions signal. Recent NLO calculations of the virtual graviton exchange cross section~\cite{K-factorII} showed that such a choice of signal $K$-factor is a reasonable assumption.  That study calculated the NLO $K$-factor to be 1.3 for masses around 500 GeV.

\begin{figure*}[tbp]
\begin{center}
\epsfxsize=7.0in
\epsffile{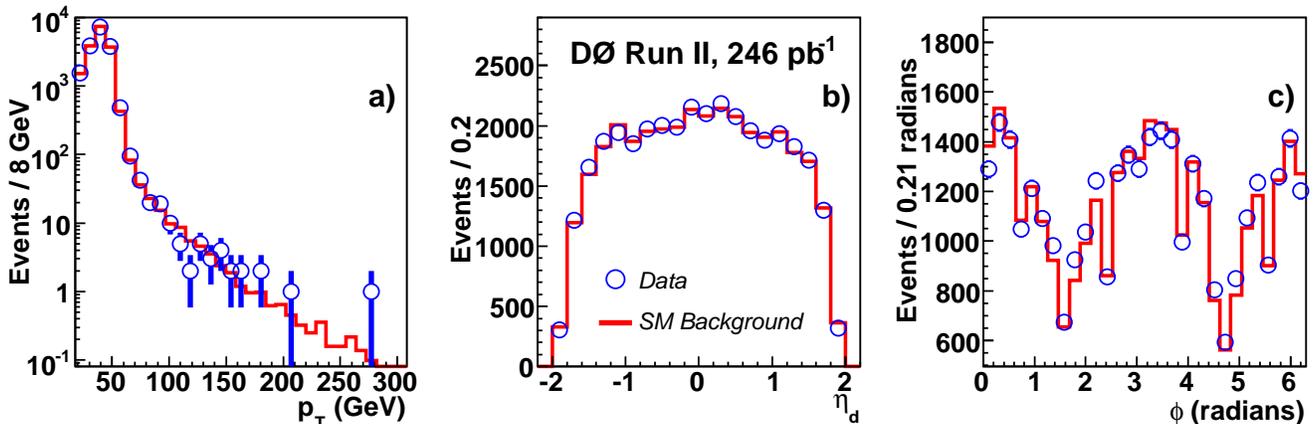} 
\vspace*{-0.2in}
\caption{
Comparison of data (circles with error bars) and SM predictions (histogram) in $p_T$, $\eta_d$, and $\phi$~\cite{phi} of the muons in the event.  The dips in the azimuthal angle reflect detector acceptance.}
\vspace*{-0.2in}
\label{fig1}
\end{center}
\end{figure*}

The main SM source of isolated dimuons is DY production, which was modeled via the MC discussed above.  Other SM sources (such as $b\bar{b}$, $Z\gamma$, $WW$, $Z \to \tau\tau$, and $t\bar t$ production) are negligible, as they either have small cross sections compared to that for DY or are eliminated by the event selection. The SM DY background prediction reproduces the main kinematic characteristics of the candidate sample, as illustrated in Fig.~\ref{fig1}. 

\begin{figure*}[bhpt]
\begin{center}
\epsfxsize=7.0in
\epsffile{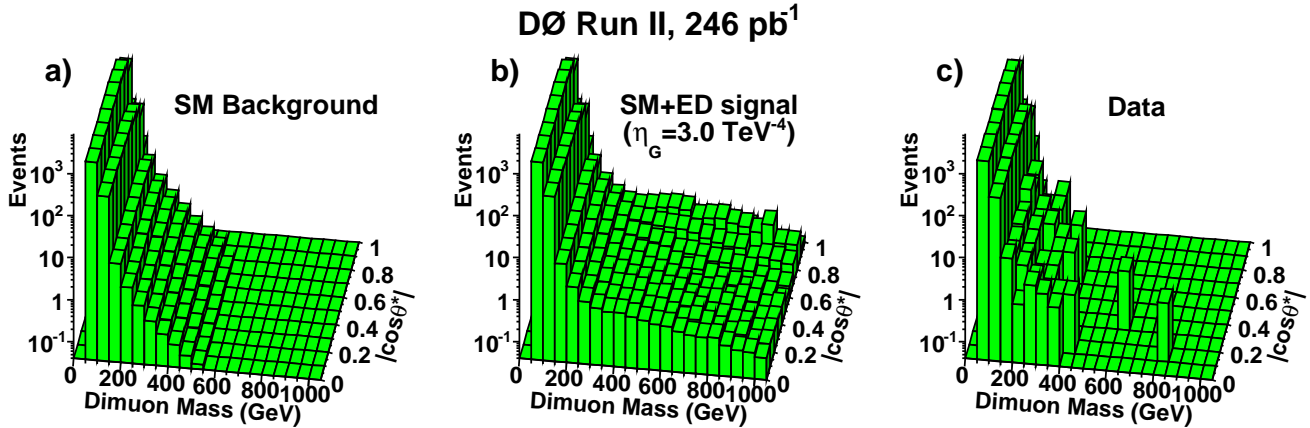} 
\vspace*{-0.2in}
\caption{
Two-dimensional distributions in the dimuon mass vs. $|\cos\theta^*|$ for: (a) SM background, (b) the sum of the SM and large extra dimensions contributions for $\eta_G = 3$~TeV$^{-4}$, and (c) data.}
\label{fig2}
\end{center}
\end{figure*}

\begin{figure}[tpb]
\begin{center}
\epsfxsize=3.3in
\epsffile{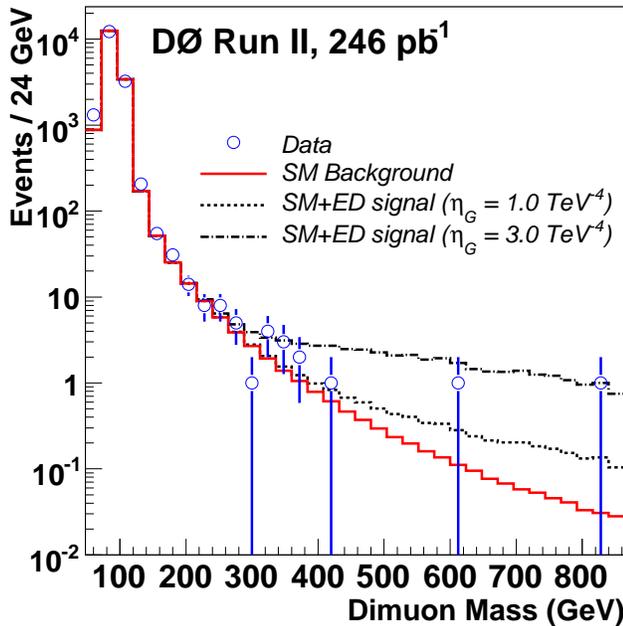} 
\vspace*{-0.2in}
\caption{Comparison between data and SM background in the dimuon mass $M$, where the effects of extra dimensions are enhanced and shown for $\eta_G = 1$~TeV$^{-4}$ and 3~TeV$^{-4}$ (dashed lines).}
\vspace*{-0.2in}
\label{fig3}
\end{center}
\end{figure}

Figure~\ref{fig2} shows the two-dimensional distribution in $M$ vs. $|\ct|$ for the SM background (other backgrounds are negligible and ignored), the sum of the background and an extra dimensions signal of $\eta_G = 3$~TeV$^{-4}$, and the data.  The data agree with the background prediction and do not exhibit evidence for large extra dimensions, which would produce an excess of events at high mass. The two highest-mass events, while having properties typical for the signal, are still in good agreement with the SM predictions alone.  To further illustrate the agreement between the SM background and data, Fig.~\ref{fig3} shows the one-dimensional mass distribution.  For reference, the background prediction is 4 events for masses greater than 400 GeV and we see 3 events in the data.

\indent We set limits on the fundamental Planck scale $M_S$ via a Bayesian fit to the data with the full signal cross section given by Eq.~(\ref{eq4}) in the entire ($M$, $|\ct|$) plane shown in Fig.~\ref{fig2}. The fit parameter $\eta_G$ was assumed to have a flat prior distribution.  Systematic uncertainties on signal and background were accounted for in the fit and include $K$-factor shape (5$\%$), the modeling of the \pt\ smearing in the MC (6$\%$), the dependence on the choice of PDF (5$\%$), $p_T$ dependence of the muon efficiency (5$\%$), and the MC-to-data normalization fit (1$\%$). The latter uncertainty accounts for the fact that we used $Z \to \mu\mu$ events in the signal sample to find the overall normalization for the MC predictions, which reduced significantly the overall uncertainty on the product of the efficiencies and the integrated luminosity.  This normalization was done in the $Z$ boson mass peak region from 50 GeV to 120 GeV.
\indent The best estimates of the parameter $\eta_G$ found in this analysis are:
\begin{eqnarray}
	\eta_G = & \hspace{2.75 mm} 0.00 & ^{\displaystyle + 0.32}_{\displaystyle - 0.00} \mbox{~TeV}^{-4}\; (\eta_G \ge 0) \label{eq5}  \\
        \eta_G = & -0.36 & \pm 0.35 \mbox{~TeV}^{-4}\; (\eta_G \le 0), \label{eq6}      
\end{eqnarray}
which are fully consistent with the SM value of $\eta_G = 0$.  From this the one-sided 95\% C.L. limits on $\eta_G$ are:
\begin{eqnarray}
	\eta_G < & \hspace{2.75 mm} 0.76&  ~\mbox{TeV}^{-4}\; (\eta_G \ge 0) \label{eq7}\\
	\eta_G >  &-0.84&  ~\mbox{TeV}^{-4}\; (\eta_G \le 0). \label{eq8}
\end{eqnarray}
Our results are in good agreement with the expected sensitivity, as obtained by an ensemble of MC trial experiments ($0.76$~TeV$^{-4}$ for $\eta_G > 0$). The use of both the mass and angular variables in the fit allowed for $\approx 7\%$ improvement in the sensitivity to $\eta_G$.

We express these results in terms of limits on the fundamental Planck scale within the three formalisms of Eqs.~(\ref{eq1})--(\ref{eq3}). In the formalism of Hewett~\cite{Hewett}, both signs of $\eta_G$ are possible and therefore both limits, (\ref{eq7}) and (\ref{eq8}), are relevant.  In the other two formalisms~\cite{GRW,HLZ}, $\eta_G$ is always positive, and only the first limit is relevant. For the HLZ formalism, the case of $n = 2$ is special since ${\cal F}$, and therefore $\eta_G$, depends on $M^2$. To relate $\eta_G$ to $M_S$ for $n=2$, we used an average $M^2$ for the $G_{\rm KK}$ term at the Tevatron of (0.64~TeV)$^2$~\cite{KCGL}.  The limits are summarized in Table~\ref{table1}.

\begin{table}[ht]
\vspace*{-0.1in}
\caption{Lower limits at the 95\% C.L. on the fundamental Planck scale, $M_S$, in TeV.}
\label{table1}
\begin{tabular}{c@{}cccccc@{}cc}
\hline
\hline
GRW & \multicolumn{6}{@{}c}{HLZ} & \multicolumn{2}{@{}c}{Hewett} \\
\hline
& ~~$n$=2 & $n$=3 & $n$=4 & $n$=5 & $n$=6 & $n$=7~~ & ~~$\lambda=+1$ & $\lambda=-1$ \\
1.07 & ~~1.09	& 1.27	  & 1.07   & 0.97   & 0.90   & 0.85~~  & ~~0.96      & 0.93 \\
\hline
\hline
\end{tabular}
\vspace*{-0.1in}
\end{table}

In summary, we have performed the first search for large extra spatial dimensions in the dimuon channel at hadron colliders by looking for effects of virtual Kaluza-Klein gravitons.  We found no evidence for large extra dimensions in this channel with $\approx 250$ \ipb\ of data collected in Run II of the Fermilab Tevatron Collider.  We set a 95\% C.L. upper limit of 0.76~TeV$^{-4}$ on the parameter $\eta_G$ (for $\eta_{G}\ge0$) that describes the strength of the extra dimensions effects.  This result corresponds to limits on the fundamental Planck scale ranging between 0.85 and 1.27~TeV for several formalisms and numbers of large extra dimensions.  For comparison in Run I D\O\ placed a limit on $M_{S}$, in dielectron plus diphoton production, of 1.1~TeV in Hewett's $\lambda=+1$ formalism, while LEP's DELPHI experiment placed a limit on $M_{S}$ in dimuon production of 0.73~TeV in Hewett's $\lambda=+1$ formalism~\cite{review2}.  The limits from this analysis represent the most restrictive achieved in the dimuon channel to date.  Results presented here also represent the most precise test of high mass SM Drell-Yan production in this channel at a hadron collider.

\input acknowledgement_paragraph_r2.tex
\end{document}

%% file: list_of_authors_r2.tex
%
\author{                                                                      
V.M.~Abazov,$^{35}$                                                           
B.~Abbott,$^{72}$                                                             
M.~Abolins,$^{63}$                                                            
B.S.~Acharya,$^{29}$                                                          
M.~Adams,$^{50}$                                                              
T.~Adams,$^{48}$                                                              
M.~Agelou,$^{18}$                                                             
J.-L.~Agram,$^{19}$                                                           
S.H.~Ahn,$^{31}$                                                              
M.~Ahsan,$^{57}$
% ...
\\                                                                            
\vskip 0.30in                                                                 
\centerline{(D\O\ Collaboration)}  
}
\affiliation{
\vskip 0.2in                                                              
\centerline{$^{1}$Universidad de Buenos Aires, Buenos Aires, Argentina}       
\centerline{$^{2}$LAFEX, Centro Brasileiro de Pesquisas F{\'\i}sicas,         
                  Rio de Janeiro, Brazil}                                     
\centerline{$^{3}$Universidade do Estado do Rio de Janeiro,                   
                  Rio de Janeiro, Brazil}                                     
\centerline{$^{4}$Instituto de F\'{\i}sica Te\'orica, Universidade            
                  Estadual Paulista, S\~ao Paulo, Brazil}                     
\centerline{$^{5}$University of Alberta, Edmonton, Alberta, Canada,           
               Simon Fraser University, Burnaby, British Columbia, Canada,}   
\centerline{York University, Toronto, Ontario, Canada, and                    
         McGill University, Montreal, Quebec, Canada}                         
\centerline{$^{6}$Institute of High Energy Physics, Beijing,                  
                  People's Republic of China}                                 
\centerline{$^{7}$University of Science and Technology of China, Hefei,       
                  People's Republic of China}                                 
\centerline{$^{8}$Universidad de los Andes, Bogot\'{a}, Colombia}             
\centerline{$^{9}$Center for Particle Physics, Charles University,            
                  Prague, Czech Republic}                                     
\centerline{$^{10}$Czech Technical University, Prague, Czech Republic}        
\centerline{$^{11}$Center for Particle Physics, Institute of Physics,         
                   Academy of Sciences of the Czech Republic,                 
                   Prague, Czech Republic}                                    
\centerline{$^{12}$Universidad San Francisco de Quito, Quito, Ecuador}        
\centerline{$^{13}$Laboratoire de Physique Corpusculaire, IN2P3-CNRS,         
                  Universit\'e Blaise Pascal, Clermont-Ferrand, France}       
\centerline{$^{14}$Laboratoire de Physique Subatomique et de Cosmologie,      
                  IN2P3-CNRS, Universite de Grenoble 1, Grenoble, France}     
\centerline{$^{15}$CPPM, IN2P3-CNRS, Universit\'e de la M\'editerran\'ee,     
                  Marseille, France}                                          
\centerline{$^{16}$IN2P3-CNRS, Laboratoire de l'Acc\'el\'erateur              
                  Lin\'eaire, Orsay, France}                                  
\centerline{$^{17}$LPNHE, IN2P3-CNRS, Universit\'es Paris VI and VII,         
                  Paris, France}                                              
\centerline{$^{18}$DAPNIA/Service de Physique des Particules, CEA, Saclay,    
                  France}                                                     
\centerline{$^{19}$IReS, IN2P3-CNRS, Universit\'e Louis Pasteur, Strasbourg,  
                France, and Universit\'e de Haute Alsace, Mulhouse, France}   
\centerline{$^{20}$Institut de Physique Nucl\'eaire de Lyon, IN2P3-CNRS,      
                   Universit\'e Claude Bernard, Villeurbanne, France}         
\centerline{$^{21}$III. Physikalisches Institut A, RWTH Aachen,               
                   Aachen, Germany}                                           
\centerline{$^{22}$Physikalisches Institut, Universit{\"a}t Bonn,             
                  Bonn, Germany}                                              
\centerline{$^{23}$Physikalisches Institut, Universit{\"a}t Freiburg,         
                  Freiburg, Germany}                                          
\centerline{$^{24}$Institut f{\"u}r Physik, Universit{\"a}t Mainz,            
                  Mainz, Germany}                                             
\centerline{$^{25}$Ludwig-Maximilians-Universit{\"a}t M{\"u}nchen,            
                   M{\"u}nchen, Germany}                                      
\centerline{$^{26}$Fachbereich Physik, University of Wuppertal,               
                   Wuppertal, Germany}                                        
\centerline{$^{27}$Panjab University, Chandigarh, India}                      
\centerline{$^{28}$Delhi University, Delhi, India}                            
\centerline{$^{29}$Tata Institute of Fundamental Research, Mumbai, India}     
\centerline{$^{30}$University College Dublin, Dublin, Ireland}                
\centerline{$^{31}$Korea Detector Laboratory, Korea University,               
                   Seoul, Korea}                                              
\centerline{$^{32}$CINVESTAV, Mexico City, Mexico}                            
\centerline{$^{33}$FOM-Institute NIKHEF and University of                     
                  Amsterdam/NIKHEF, Amsterdam, The Netherlands}               
\centerline{$^{34}$Radboud University Nijmegen/NIKHEF, Nijmegen, The          
                  Netherlands}                                                
\centerline{$^{35}$Joint Institute for Nuclear Research, Dubna, Russia}       
\centerline{$^{36}$Institute for Theoretical and Experimental Physics,        
                  Moscow, Russia}                                             
\centerline{$^{37}$Moscow State University, Moscow, Russia}                   
\centerline{$^{38}$Institute for High Energy Physics, Protvino, Russia}       
\centerline{$^{39}$Petersburg Nuclear Physics Institute,                      
                   St. Petersburg, Russia}                                    
\centerline{$^{40}$Lund University, Lund, Sweden, Royal Institute of          
                   Technology and Stockholm University, Stockholm,            
                   Sweden, and}                                               
\centerline{Uppsala University, Uppsala, Sweden}                              
\centerline{$^{41}$Lancaster University, Lancaster, United Kingdom}           
\centerline{$^{42}$Imperial College, London, United Kingdom}                  
\centerline{$^{43}$University of Manchester, Manchester, United Kingdom}      
\centerline{$^{44}$University of Arizona, Tucson, Arizona 85721, USA}         
\centerline{$^{45}$Lawrence Berkeley National Laboratory and University of    
                  California, Berkeley, California 94720, USA}                
\centerline{$^{46}$California State University, Fresno, California 93740, USA}
\centerline{$^{47}$University of California, Riverside, California 92521, USA}
\centerline{$^{48}$Florida State University, Tallahassee, Florida 32306, USA} 
\centerline{$^{49}$Fermi National Accelerator Laboratory, Batavia,            
                   Illinois 60510, USA}                                       
\centerline{$^{50}$University of Illinois at Chicago, Chicago,                
                   Illinois 60607, USA}                                       
\centerline{$^{51}$Northern Illinois University, DeKalb, Illinois 60115, USA} 
\centerline{$^{52}$Northwestern University, Evanston, Illinois 60208, USA}    
\centerline{$^{53}$Indiana University, Bloomington, Indiana 47405, USA}       
\centerline{$^{54}$University of Notre Dame, Notre Dame, Indiana 46556, USA}  
\centerline{$^{55}$Iowa State University, Ames, Iowa 50011, USA}              
\centerline{$^{56}$University of Kansas, Lawrence, Kansas 66045, USA}         
\centerline{$^{57}$Kansas State University, Manhattan, Kansas 66506, USA}     
\centerline{$^{58}$Louisiana Tech University, Ruston, Louisiana 71272, USA}   
\centerline{$^{59}$University of Maryland, College Park, Maryland 20742, USA} 
\centerline{$^{60}$Boston University, Boston, Massachusetts 02215, USA}       
\centerline{$^{61}$Northeastern University, Boston, Massachusetts 02115, USA} 
\centerline{$^{62}$University of Michigan, Ann Arbor, Michigan 48109, USA}    
\centerline{$^{63}$Michigan State University, East Lansing, Michigan 48824,   
                   USA}                                                       
\centerline{$^{64}$University of Mississippi, University, Mississippi 38677,  
                   USA}                                                       
\centerline{$^{65}$University of Nebraska, Lincoln, Nebraska 68588, USA}      
\centerline{$^{66}$Princeton University, Princeton, New Jersey 08544, USA}    
\centerline{$^{67}$Columbia University, New York, New York 10027, USA}        
\centerline{$^{68}$University of Rochester, Rochester, New York 14627, USA}   
\centerline{$^{69}$State University of New York, Stony Brook,                 
                   New York 11794, USA}                                       
\centerline{$^{70}$Brookhaven National Laboratory, Upton, New York 11973, USA}
\centerline{$^{71}$Langston University, Langston, Oklahoma 73050, USA}        
\centerline{$^{72}$University of Oklahoma, Norman, Oklahoma 73019, USA}       
\centerline{$^{73}$Brown University, Providence, Rhode Island 02912, USA}     
\centerline{$^{74}$University of Texas, Arlington, Texas 76019, USA}          
\centerline{$^{75}$Southern Methodist University, Dallas, Texas 75275, USA}   
\centerline{$^{76}$Rice University, Houston, Texas 77005, USA}                
\centerline{$^{77}$University of Virginia, Charlottesville, Virginia 22901,   
                   USA}                                                       
\centerline{$^{78}$University of Washington, Seattle, Washington 98195, USA}  
}                                                                             

%% file: acknowledgement_paragraph_r2.tex
%
We thank the staffs at Fermilab and collaborating institutions, 
and acknowledge support from the 
DOE and NSF (USA);
CEA and CNRS/IN2P3 (France);
FASI, Rosatom and RFBR (Russia);
CAPES, CNPq, FAPERJ, FAPESP and FUNDUNESP (Brazil);
DAE and DST (India);
Colciencias (Colombia);
CONACyT (Mexico);
KRF (Korea);
CONICET and UBACyT (Argentina);
FOM (The Netherlands);
PPARC (United Kingdom);
MSMT (Czech Republic);
CRC Program, CFI, NSERC and WestGrid Project (Canada);
BMBF and DFG (Germany);
SFI (Ireland);
Research Corporation,
Alexander von Humboldt Foundation,
and the Marie Curie Program.

%% file: PRL_LED_final.bbl
\begin{thebibliography}{}
\bibitem{phi}
We use a right-handed coordinate system with the z-axis in the direction of the proton beam and azimuthal angle $\phi$ defined in the plane perpendicular to the beams with $\phi = \pi/2$ pointing up.
\end{thebibliography}
